\documentclass{elsart5p}

\usepackage{amssymb}

\newcommand{\be}{\begin{equation}}
\newcommand{\ee}{\end{equation}}
\newcommand{\ba}{\begin{array}}
\newcommand{\ea}{\end{array}}
\newcommand{\bk}{{\bf k}}

\begin{document}

\begin{frontmatter}

\title{ 
Stability of  Nagaoka phase, spin effective action and delocalized free holes}
\author[cifmc]{F. L. Braghin}
\address[cifmc]{
International Center for Condensed Matter Physics -UnB,
C.P. 04513, 70.904-970, Brasilia, DF, Brazil
}

\begin{abstract}
The Hubbard model in the limit of infinite $U$ is investigated
within a projected slave fermion representation
and following a previous work of the author and collaborators \cite{BFK}.
The stability of the Nagaoka's phase with respect to 
a non vanishing concentration of holes ($\delta_h$) is analyzed by envisaging the existence of a spin effective action for itinerant magnetism of the Hubbard model. 
It is considered that,
as the hole doping increases away from the half filled insulating
limit, free holes are expected to be more delocalized.
Depending on treatment for the hopping: a ferromagnetic or anti-ferromagnetic 
ordering  might arise  and
the Nagaoka
phase might  have some stability with respect
to $\delta_h \neq 0$.
\end{abstract}


\end{frontmatter}

\section{ Introduction }

One of the few exact
 results for the Hubbard model (HM)
is the ferromagnetic Nagaoka limit for $U=\infty$ \cite{nagaoka,tasaki}. 
This phase appears when
one hole hops in the half filled band.
Although this limit is not found in any material, it can be reached in 
optical traps \cite{optical}.
The investigation of the Nagaoka mechanism 
provides relevant information about the phase diagram of the model
and one starting point for understanding further the role of 
strong electronic correlations 
 \cite{BFK,fradkin-auerbach,wiegmann,anderson}
and of  realistic
mechanism for itinerant (ferro)magnetism
 \cite{fradkin-auerbach,tasaki}. 
In the limit of infinite Coulomb repulsion the HM is written as:
$$
H=-\sum_{ij,\sigma}t_{ij}\tilde c^{\dagger}_{i\sigma}\tilde c_{j\sigma}+\mu\sum_{i\sigma}(1-\tilde c^{\dagger}_{i\sigma}
\tilde c_{i\sigma}),
$$
where $t_{ij}$ is a symmetric matrix with elements representing
 the hopping amplitude $t$ only non-zero between the
nearest-neighbor sites; $\tilde c_{i\sigma}$ is the projected electronic
operator \cite{fradkin-auerbach}.
In this expression the chemical potential $\mu$ is to control
the number of vacancies (away from half filling), and the projected electronic operator 
carries the effect of the strong correlations, i.e. it
excludes the doubly occupied states.

It is difficult to handle the strong electronic correlations
and thus
to provide exact results,
in particular 
concerning the stability of the Nagaoka phase. 
However, a common trend is that
this FM phase is unstable with respect 
to a finite concentration of holes in particular in the thermodynamic limit
 \cite{shastry-etal-90,suto,park-etal-2007,OPK-97,zitzler-etal-04,LONG-rasetti-94,c-p}.
In the present work we investigate 
the  role of the delocalization of (free) holes 
for the appearance and for the stability of the Nagaoka's phase
following the long-wavelength approach with slave fermion representation
 worked out in
Ref.~\cite{BFK}.
By envisaging the derivation of a spin-effective action ($S_{eff}$) 
for the HM with very large $U$,
a previous analysis was performed in Ref.~\cite{BFK}.
On the other hand, in the present work, the role of 
delocalization of free holes close to half filling is investigated. 
It is considered that 
the increase of the number or concentration of holes should increase
 the mobility of holes departing from the half filled limit.
An itinerant magnetic phase (ferromagnetic or anti-ferromagnetic) emerges
 depending on the structure and treatment of the hopping
of spinless holes.

\section{Slave fermion for the $U=\infty$ Hubbard model}

To account for the strong correlations that forbid  doubly occupied states, 
consider the slave-fermion decomposition for projected 
electronic operators \cite{sf} given by:
$
\tilde{c}_{i \sigma}^{\dagger} = 
b_{i \sigma}^{\dagger} f_i,$
where two operators have been used: 
$b_{i \sigma}$ stands for a spinon boson  and $f^{\dagger}_i$
creates a charged spinless fermion. 
The functional generator for the $U=\infty$ Hubbard model is given by:
$ Z_{U=\infty} = \int {\cal D}
[b, b^{\dagger};f,f^{\dagger}] 
 \exp\left( 
S_{U=\infty}^{SF} [b_i, b^{\dagger}_i;f_i,f^{\dagger}_i]
 \right)$.
The action can be written,
with the time-dependent phase,
as
\cite{BFK}:
\begin{eqnarray} 
S_{U=\infty}^{SF} & =& - \int d \tau \sum_{<i,j>}
f_i \left[ \left( \partial_{\tau} + \mu \right) \delta_{ij}
+ \sum_{\sigma} t_{ij} b^{\dagger}_{i \sigma} b_{j \sigma} 
\right]  f^{\dagger}_j 
\nonumber
\\
&& - \int d\tau \sum_{i}\ \left( \sum_{\sigma}\ b^{\dagger}_{i \sigma} 
\partial_{\tau} 
b_{i \sigma} - H_{constr} \right)
\label{S-SF}
\end{eqnarray}
The local non doubly occupancy (NDO) constraint is imposed by a (local) 
Lagrange multiplier,
$\lambda_i$,  by adding the term:
$
H_{constr} = \lambda_i ( f_i^{\dagger} f_i + 
\sum_{\sigma = \uparrow, \downarrow}\
b_{i \sigma}^{\dagger} b_{i \sigma}  - 1)
$.

The slave-fermion decomposition is equivalent to a particular (lowest weight)
representation of the $su(2|1)$ super-symmetric  
 projected electronic operators \cite{wiegmann,FKM}:
 spinless holes are super-partners of  spinons. 
A
mapping for the variables,  incorporating implicitly the non-doubly-occupancy 
(NDO) constraint, is given by \cite{BFK}:
\begin{eqnarray}
( b_{i\uparrow }, \;\;\; b_{i\downarrow }, \;\;\; f_i ) = 
  \frac{ ( e^{i\phi_i}, \;\;\; z_ie^{i\phi_i}, \;\;\; \xi_i
 e^{i\phi_i} )
}{\sqrt{1+\overline{z}_iz_i
+\overline{\xi}_i\xi_i}}, 
\label{mapping}
\end{eqnarray}
and the corresponding variables for $b_{i,\sigma}^{\dagger}$ and $f_i^{\dagger}$.
With these new variables ($z_i, \xi_i, \phi_i$ respectively for bosonic 
spinons, spinless fermions and a local phase) the
Lagrange multiplier $\lambda_i$ is eliminated naturally, being the 
NDO constraint incorporated. 
With the spinon variables $z_i$, 
the images of the spin $su(2)$ algebra - $\vec{S}$ - can be rewritten 
 \cite{BFK,FKM},
for example: 
$S_z^{cl}=\frac{1}{2}\frac{1-|z|^2}{1+|z|^2}$. 
There is a local gauge invariance as consequence of the redundancy in parameterization of 
the electron operator in terms
of the auxiliary boson/fermion fields. 
After some manipulation,
 decomposing the measure of the path integral into the new variables
${\cal D}
[b, b^{\dagger};f,f^{\dagger}] 
 \rightarrow D\mu_{spin}(\overline{z},z)\times D\mu_{fermion}(
\overline{\xi },\xi )$, with the corresponding Jacobian \cite{FKM}, 
the action
reads \cite{FKM,BFK}:
\begin{eqnarray}
S &=& \sum_i \int_0^{\beta} i a_i (\tau) d\tau- \sum_i\ \int_0^{\beta}
\bar\xi_i
\left(\partial_{\tau} + \mu + ia_i \right)
\xi_i d \tau 
\nonumber
\\
&& -
\int_0^{\beta}   t \sum_{ij}(\overline{\xi }_{j}\xi _{i} 
< z_{i}|z_{j} > +h c )    \, d\tau 
\label{S-su21}
\end{eqnarray}
The first
term of this action is a  kinematical term 
and the second is the
classical image of the Hamiltonian.
The spin
"kinetic" term
(Berry phase):
$ia=- < z|\partial_t|z > =\frac{1}{2} 
\frac{\dot{\bar z}z-\bar z\dot
z}{1+|z|^2},$ with $|z\rangle$ being the su(2) coherent state \cite{FKM,BFK}.
The inner product of the su(2)
coherent states is written as:
 $\langle z_{i}|z_{j}\rangle  = \frac{1+\overline{z}
_{i}z_{j}}{\sqrt{(1+|z_{j}|^{2})(1+|z_{i}|^{2})}} \equiv \frac{1}{t_{ij}}\Sigma_{ij}$.

\section{  Factorization of the hopping}

By introducing more holes in the half filled HM, it can be expected 
they become
progressively more delocalized.
Consider that the band structure is such that the hopping term can be 
decomposed  into two parts.
One of them endows the holes with a dispersion relation (labeled by $\gamma_1$)
and  the 
other is treated as a perturbation 
 (labeled by $\gamma_2$), 
eventually from a different band.
It will be considered schematically 
that:
\begin{eqnarray} \label{g0-sf-0}
\bar{\xi}_i \Sigma_{ij} \xi_j
\to \gamma_1 \; \bar{\xi}_j^{(1)} \xi_i^{(1)} \Sigma_{ij}^{(1)} + 
\gamma_2 \;
\bar{\xi}_j^{(2)} \xi_i^{(2)}
\Sigma_{ij}^{(2)} + h.c.
\end{eqnarray}
Where $\gamma_1$ and $\gamma_2$  keep track 
of each of the different parts of the hopping.
The procedure and idea will be clearer and useful when working in 
the momentum space.
We will consider that these terms are characterized by different ranges of
momenta of holes  $\xi^{(1)}(\bk_1)$ and $\xi^{(2)}(\bk_2)$, 
associated respectively to the terms $\Sigma^{(1)}$ and $\Sigma^{(2)} $, 
such that  $\bk_1$ and  $\bk_2$ can belong to different parts of the band.
%eventually low and high energy modes.
With this decomposition,
the following  ansatz for the free hole Green's 
functions  can be envisaged:
\begin{eqnarray} \label{g0-sf-1}
(G_0^{-1})_{ij} = \left( \partial_{\tau} - \mu
 \right) \delta_{ij} \delta (\tau) 
+ \gamma_1 {\Sigma^{(1)}}_{ij} 
,
\end{eqnarray}
and  $\gamma_2 .\Sigma^{(2)} = \gamma_2 t_{ij}^{(2)} <z_i|z_j>^{(2)}$
is a perturbation.
The upper indices $^{(2)}$ stand for the perturbation
due to the corresponding part of the 
hopping term, separated according to expression (\ref{g0-sf-0}).
Since this separation is generic and not calculated microscopicaly,
for the sake of generality we can have $t^{(2)}_{ij} \neq t^{(1)}_{ij}$
depending on the hopping (and band) structure.
This procedure can be considered such as to 
provide a measure of the (de)localization of the free holes. 
For instance, in a normal conducting phase, we should recover 
$\gamma_2 \to 0$, that is used in the usual mean field approximation
\cite{boies-etal-95,BFK}.
On the other hand, at the half filling limit (and very close to it)
 the hopping parameter would 
be such that
$\gamma_1 = 0$, suitable for the hopping (loop) expansion 
as discussed in details in Ref.~\cite{BFK}.

\subsection{Delocalization of free holes and spin effective action}

 We take  a continuum limit of the 
full action, given by 
(\ref{S-su21}), to derive a spin-effective action, $S_{eff}$,
by integrating out the fermion variables with the prescription
(\ref{g0-sf-1})
in the same lines as it was done in Ref.~\cite{BFK}.
Using a finite difference method for the term labeled by $\gamma_1$,
we take:
$\xi_{j=i+1} \to \tilde{\xi} (i) + a \nabla \tilde{\xi} (i)$
and perform a Fourier transformation.
Therefore  we consider 
 free holes are endowed with a dispersion relation $\epsilon (\bk_{(1)})$,
being that $\bk_1$ ($\bk_2$)refers to the momenta of modes labeled by
$\gamma_1 (\gamma_2)$.
This yields the momentum dependent Green's function:
$G_0 [\mu; \epsilon({\bf k}^{(1)}) ]$.
The particular dispersion relation $\epsilon(\bk^{(1)})$ is  completely defined by 
the lattice (geometry and dimensionality).
For the sake of the main argument,
we consider a two dimensional square lattice,
for which it follows:
$\epsilon ({\bf k_{(1)}}) 
\simeq 2 \gamma_{(1)} t \sum_{k,\sigma}\ 
\phi_{k,\sigma}^2 \left( cos (k_x) + cos (k_y)
\right), 
$ from what the continuum limit is extracted.
Prescription (\ref{g0-sf-0})
might also be associated
to a superposition of (nearly) localized and (fully)
 delocalized states.
In order simplify the notation
$\bk_{(1)}$ momenta will
 be denoted simply by $\bk$ from here on.

The corresponding effective action, with the integration of 
fermions variables, can be written as \cite{BFK}:
\begin{eqnarray} 
S_{eff} = Tr \mbox{Log} G^{-1} \equiv Tr \mbox{Log} 
\left( G_0^{-1} - i a + \Sigma^{(2)} \right) 
\nonumber
\\
= Tr \mbox{Log} G^{-1}_{0}
+
 Tr \mbox{Log} ( 1 - G_{0} ia + G_{0}\Sigma^{(2)} ).
\label{S-eff-exact}
\end{eqnarray}
The different modes of the fermions
are decoupled such that the corresponding
 $\Sigma^{(i)}$ are treated (nearly) independently.
The free hole Green's function can be calculated, for the sake of generality, 
for a given (sub)lattice $A$, instead of an unique lattice, 
being written as:
 $(G^{-1}_0)^A (\mu^A, {\bf k}) = 
(\partial_{\tau} - \mu^A + \epsilon_A({\bf k}))^{-1}
\delta( \tau)$. 
This case of (at least)
two sublattices will not be worked out here, and this might be considered
when there are different hoppings in the sublattices or between each of 
them.  
This can yield the terms labeled by $\gamma_1$ and $\gamma_2$
contributing in each of the different sublattice.
The long-wavelength expansion is done by considering 
$G_0^{-1} > > \Sigma^{(2)}$. 
The reliability of this 
expansion depends on several parameters, seen in
expressions (\ref{g0-sf-0}) and (\ref{g0-sf-1}).
Basically it is required that 
$\mu + \epsilon(\bk) >  > \Sigma^{(2)}$
where $\Sigma^{(2)}$ is only part of the full hopping term.
We remind further that $\Sigma^{(2)}$
is basically proportional to  $t$ 
and the long-wavelength limit corresponds to a gradient expansion of
 $t <z_i|z_j>$. 
Therefore  we expect to provide
a complementary investigation to the loop expansion analyzed in Ref.~\cite{BFK}.
considering the role of the delocalization of free holes.
For the sake of the argument and to show preliminary analytical results,
 $\Sigma^{(1)}$ and $\Sigma^{(2)}$ are considered to somehow
decouple from each other.
In this case the expression for the leading terms of the effective action,
is obtained in the very same way as shown in Ref.\cite{BFK}. 
Keeping track of the time ordering in the path integral with 
a Taylor expansion in fluctuating times \cite{CR,BFK}, for D-dimensions
$S_{eff}$ 
is given by:
\be \ba{ll} \label{free-energy-1}
\displaystyle{ {\cal S}_{eff} =  
\int \frac{d^D \bk}{(2 \pi)^{D}} \mbox{Log} 
\left( 1 + exp \left( - \beta ( \mu  - \epsilon (\bk) )\right) \right)
} \\
\displaystyle{ 
- \sum_{<i,j>}\ \int_0^{\beta}  \frac{J_{eff} }{2}
| <z_i|z_j> |^2  d \eta  
- 
 \sum_i
\int_0^{\beta} d \eta K_{eff} \; i \;a_0 (\eta)   +  
...}
\ea \ee
where $...$ stands for the higher order terms.
In the case the spinon dynamics decouples completely from the holes, 
we rewrite $S_{eff}$ in 
 the momentum space, with  zero momentum transfer
 between holes and spinons. 
The effective coefficients $J_{eff}$ and $K_{eff}$
can be written as:
\begin{eqnarray}
J_{eff} = - \gamma_2^2 \int_{K_0}^{K_1} \frac{d^D \bk}{(2 \pi)^{D}} 
\frac{ t^2 \beta}{4 cosh^2 \left( \frac{\beta (\mu - \epsilon(\bk))}{2}\right)}
\\
K_{eff} = - \int_{K_0}^{K_1} \frac{d^D \bk}{(2 \pi)^{D}} \frac{2
}{
\left( e^{\left( \beta (\mu - \epsilon (\bk) )\right)}  + 1 \right)
}.
\end{eqnarray}
The $K_{eff}$ is the coefficient of the time dependent term and it 
will not be analyzed here, since it does not modify the magnetic ordering
in a first analysis.
$K_0$ and $K_1$ are the (upper and lower) limiting values of the momenta of holes 
which contribute in that part of the decomposition of the variables,
i.e. holes from $\gamma_1$ term.
For $\gamma_1 = 0$, the expressions obtained in Ref.\cite{BFK} are reproduced.
In this expression the second order (leading) term can be written 
in the form of a Heisenberg coupling (either in k-space or in the lattice)
as:
$$H^{cl}_{eff}=  \frac{J_{eff}}{2}  \sum_{(ij)}|\langle z_i|z_j\rangle|^2 =
J_{eff} \sum_{(ij)}  (\vec S_i\vec S_j+ \frac{1}{4} ),$$
where the  classical symbols of the spin operators for
the quantum $s=1/2$  Heisenberg model were used  \cite{BFK}, 
and the corresponding modes $\bk_1$ and $\bk_2$ have been separated. 
Below we show some expressions for the coefficient of the 
Heisenberg  spin-coupling
$J_{eff}$.
The second order term in the effective action (\ref{free-energy-1})
has different signs and structures
 for $J_{eff}$ depending on the range of the parameters. 
However it is worth emphasizing that due to this separation, 
the spinon connection ($<z_i|z_j>$) in expression (\ref{free-energy-1})
corresponds to only part of the 
full spinon dynamics (that from $\Sigma^{(2)}$), and therefore the resulting phase
might not have a  fully saturated (ferro)magnetic ordering.
The separation of the full  hopping in ranges of momenta,
labeled by $\gamma_{1}, \gamma_{2}$,
is a relevant assumption for this analysis.
A cutoff in the momentum integration might correspond to the emergence of a 
kind of Fermi surface for the vacancies in which an
integration in $k$ is limited by $k_F$, the momentum of holes at the 
(eventual) Fermi surface. 
A decomposition
 in low and high energy modes suggests a renormalization group analysis, which
will be presented elsewhere.
However separation of modes has been also
implemented in, for example, Ref.~\cite{falb-muramatsu}.

\subsection{Some analytical expressions for $J_{eff}$}

The particular dispersion relation for the holes was not explicitly written 
so far. 
For the sake of generality, two cases are considered of the following
usual forms:\\
 $\epsilon_{(I)} (\bk) = \gamma_1 \tilde{a}_1 k \equiv {a}_1 k $  and
$\epsilon_{(II)} (\bk) = \gamma_1  \tilde{b}_1 \bk^2 \equiv b_1 \bk^2$.

Changing variables for each of the cases,
we write: 
\begin{eqnarray}
J_{eff}^{(I)} &=&  \gamma_2^2 
\frac{ 2 t^2 \Omega_D \beta }{ 4 (2 \pi)^{D} \beta a_1}
\int^{X_1^{(I)}}_{X_0^{(I)}} \, d \, x \;
\frac{ \left[ \left( \mu - \frac{2 x}{\beta} \right) \frac{1}{a_1}
\right]^{D-1}
}{cosh^2(x)}
\\
J_{eff}^{(II)} &=&  \gamma_2^2 \frac{2 t^2 \Omega_D \beta}{4 (2 \pi)^D b_1 \beta }
\int^{X_1^{(II)}}_{X_0^{II}} \, d \, x \; x
\frac{ \left[ \left( \mu - \frac{2 x^2}{\beta} \right) \frac{1}{b_1}
\right]^{D-2}
}{cosh^2(x)},
\end{eqnarray}
where $\Omega_D$ is the integral of the 
$D$ dimensional solid angle; and the cutoffs are:
$\; X_0^{(I)} = \frac{(\mu - a_1 K_0) \beta}{2}$, 
$ \; X_1^{(I)} = \frac{(\mu - a_1 K_1) \beta}{2}$,
and
$\; X_0^{(II)} = \frac{(\mu - b_1 K_0^2) \beta}{2}$, 
$\; X_1^{(II)} = \frac{(\mu - b_1 K_1^2) \beta}{2}$.
With the eventual formation of a Fermi surface for 
the (spinless) vacancies in a normal metallic phase, 
we could identify the chemical potential
to: $\mu = \epsilon (\bk_F)$, where $\bk_F$ is the momentum at the Fermi
surface.
The result for the quadratic $\epsilon_{(II)}$ in 2-dim is 
the same as for $\epsilon_{(I)}$ in 1-dim, apart from a normalization.

The integrations, in 2 dim,
yield respectively:
\begin{eqnarray}
J_{eff}^{(I), D=2} & =& - \gamma_2^2 \frac{t^2}{4 \pi a_1 \beta}
\left[ X_1^{(I)} tanh ( X_1^{(I)} ) 
- X_0^{(I)} tanh ( X_0^{(I)} ) 
 \right. 
\nonumber
\\
& &\left.
+ \frac{2}{\beta} \mbox{Log} 
\left( \frac{cosh X_1^{(I)} }{cosh X_0^{(I)} } \right)
\right]
\label{J-eff-1}
\\
J_{eff}^{(II), D=2} &=&   \gamma_2^2 \frac{t^2}{ 4 \pi b_1}
\left[ 
 tanh 
\left( X_1^{(II)}  \right) - 
tanh ( X_0^{(II)})
\right]
\label{J-eff-2}
\end{eqnarray}

A short example can be taken, for $T=0$,
 by choosing $K_1 > K_0 \simeq k_F \simeq \mu/a_1$ 
for the first of these 
expressions, in which case $K_0$ might be the momentum at a Fermi surface for the 
holes.
The resulting spin-effective coupling at zero temperature is given by:
$J_{eff}^{(I)} \simeq \gamma^2_2 t^2 ( 2 a_1 K_1 -\mu)/ (4 \pi a_1) > 0.$
This is a ferromagnetic coupling representing a Nagaoka-type phase, with
some stability since it remains finite for a finite corresponding 
concentration of holes $\delta_h > 0$.
In particular, for very small $\delta_h$, we have the compressibility of holes,
$\kappa_h (T \to 0) \to 0$.
It is interesting to notice that $\delta_h$ and $\kappa_h$ are 
calculated analytically.
As temperature increases 
the decay to a paramagnetic phase should take place,
depending strongly on 
the spinon dynamics whose investigation is outside the 
scope of the present work.
We emphasize that the present work only aim to provide a different starting point 
for investigating the role of delocalization of holes. 
The appearance of the ferromagnetic coupling however was 
 related to the range of integration of the momenta of holes.
Should we consider a different physical picture in which 
the relation among the variables $K_0,K_1, \mu$ were related 
differently, it can give rise to an 
(itinerant) anti-ferromagnetic coupling.
This analysis
remains valid for  $U < \infty$.

In the limit of no limitation on the momenta of holes,
i.e. $K_0 = 0$ and $K_1 \to \infty$,
in 2 dim, it yields respectively:
\begin{eqnarray}
J_{eff}^{I, D=2} & \to &  - \gamma_2^2 t^2
\frac{\mu}{a_1^2 \sqrt{\pi}} \sum_{k=0}^{\infty} 
\frac{(-1)^k}{\sqrt{2 k + 1}} - \frac{2 }{\pi a_1^2 \beta} (Ln 2)
\\
J_{eff}^{II, D=2} & \to & - \gamma_2^2 \frac{t^2}{ \pi b_1} \sum_{k=0}^{\infty} 
\frac{(-1)^k}{\sqrt{2 k + 1}}. 
\end{eqnarray}
We notice that these couplings might depend on $\mu$.
They provide anti-ferromagnetic spin Heisenberg couplings.

\section{Final remarks}

We have shown that the delocalization of free holes might be a relevant issue
for the understanding of the stability of the Nagaoka's phase
with respect to a finite concentration of holes. 
More generally we proposed a framework 
for investigating
different magnetic orderings in the limit of very large 
Coulomb repulsion and low concentration of holes for the Hubbard model.
A spin effective action was found to have
 a form of localized Heisenberg coupling
in the long wavelength limit along with the work
presented in Ref.~\cite{BFK}. It can be
ferromagnetic (Nagaoka-type phase) or anti-ferromagnetic
depending on the relation among the chemical potential and the 
eventual values of the limitation on the summation/integration
of momenta carried by the holes. 
For that, the hopping term was separated in two parts, corresponding to
high and low momentum modes or to hopping among different bands
\footnote{
Eventually the  parameters $\gamma_i$ introduced to label 
$\Sigma^{(i)}$
might be expected to depend on the temperature, $U$ and concentration 
of holes $\delta_h$ 
\cite{ostlund-granath,localization} becoming phenomenological.
For example, one might want to account the variation of $\gamma_{i}$
due to a finite value for the Coulomb repulsion
by means of prescriptions. They might be given by:
\begin{eqnarray} \label{ansatz1}
& (i)& \;\;\; \gamma_1 = \frac{\alpha_0}{\alpha_0 + \alpha_U U},
 \;\;\;\;\;
 \gamma_2 = \frac{\alpha_U \; U}{\alpha_0 + \alpha_U U},
\nonumber
 \\
& (ii)& \;\;\; 
\gamma_1 = \frac{ 2 \alpha_1+ \alpha_2 \; U }{2 (\alpha_1 + \alpha_U U)},
 \;\;\;\;\;
 \gamma_2 = \frac{\alpha_3 \; U}{2 (\alpha_1 + \alpha_U U)} ,
 \end{eqnarray}
 where $\alpha_i$ ($i=0,U,1,2$) 
depend on the parameters of the model and their values are bounded by
 $0$ and $1$.
 The values of such parameters must be constrained due to 
 expression (\ref{g0-sf-0}). For example
 for this second parameterization, $\alpha_2 + \alpha_3 = 2\alpha_U$.
This makes possible to consider that the hopping term
 contributes both in 
 $\Sigma$ and in $G_0$
, whereas in the first
prescription (i) we recover the development of Ref.~\cite{BFK}, 
for which $\gamma_1=0$ when $U=\infty$.
}.
A  microscopic derivation of the prescriptions adopted was not yet 
presented and 
for the sake of the main argument it was 
considered that the holes are
reasonably decoupled from the spinons.
In particular, by endowing holes with  a dispersion relation
such that a kind of Fermi surface can be formed, it was found  that
the Nagaoka's phase at finite concentration of holes
can have some  stability  in a long-wavelength limit.
A more complete analysis will be presented elsewhere.

\section*{Acknowledgements}

This work was supported by IBEM, CNPq, Ministry of Science and Technology
of Brazil. F.L.B. thanks  E.Kochetov and A. Ferraz 
for a collaboration.


\begin{thebibliography}{00}

\bibitem{BFK} F.L. Braghin, A. Ferraz, E. Kochetov,  Phys. Rev. {\bf B 78}, 
115109 (2008);
arXiv:cond-mat/0712.3431v2..

\bibitem{nagaoka}  Y. Nagaoka, Phys. Rev. 147, 392 (1966).
% Similar investigation 
% was done by: J. Thouless, Proc. Phys. Soc. London {\bf 86}, 
% 893 (1965).

\bibitem{tasaki} Hal Tasaki, Prog. Theor. Phys. 99;
cond mat/9712219v3.

\bibitem{optical} D. Jaksch and P. Zoller, Ann. Phys. {\bf 315}, 52 (2005).


\bibitem{fradkin-auerbach} A. Auerbach, Interacting Electrons and Quantum Magnetism, Springer, 
1994. E. Fradkin, Field Theories of Condensed Matter Systems, Addison Wesley, (1991).

\bibitem{wiegmann} P.B. Wiegmann, Phys. Rev. Lett. {\bf  60}, 821 (1988).

\bibitem{anderson} P.W. Anderson, Phys. Rev. Lett. {\bf  64}, 1839 (1990).

\bibitem{shastry-etal-90} B.S. Shastry {\it et al}, Phys. Rev. {\bf B 41}, 
2375 (1990).

\bibitem{suto} A. Suto, Commun. Math. Phys. {\bf 140}, 43 (1991).

\bibitem{park-etal-2007} H.Park, K. Haule, C.A. Marianetti, G. Kotliar,
arXiv:cond-mat/0708.4240; Phys. Rev. {\bf B 77}, 035107 (2008).


\bibitem{OPK-97} T. Obermeier, T. Pruschke, J. Keller, Phys. Rev. B 56,
R8479 (1997).

\bibitem{zitzler-etal-04} R. Zitzler, Th. Pruschke, R. Bulla, Journ. of 
Magnetic Materials {\bf 272}, 21 (2004).

\bibitem{LONG-rasetti-94} M.W. Long, in "The Hubbard Model, recent results",
ed. by M. Rasetti, World Scientific, (1991).

\bibitem{c-p}  P. Coleman, C. Pepin, Physica  B {\bf 312}, 539 (2002).

\bibitem{boies-etal-95} 
D. Boies, F. A. Jackson and A-M. S. Tremblay, 
Int. Jour. Mod. Phys B 9, 1001
(1995).

\bibitem{sf} D. Yoshioka, J. Phys. Soc. Jpn. {\bf 58}, 1516 (1989);
D.P Arovas and A. Auerbach, Phys. Rev. B {\bf 38}, 316 (1988).

%\bibitem{nayak} Chetan Nayak, Phys. Rev. Lett. {\bf 85}, 178 (2000).

\bibitem{FKM} A. Ferraz, E. Kochetov, 
M. Mierzejewski, Phys. Rev. {\bf B 73}, 064516 (2006).

\bibitem{CR} M. Cuoco and J. Ranninger, Phys. Rev. B {\bf 70}, 104509 (2004).

\bibitem{falb-muramatsu} J. Falb, A. Muramatsu, arXiv:cond-mat/0705.1918.

\bibitem{ostlund-granath} S. \"Ostlund, M. Granath, Phys. Rev. Lett.
{\bf 96} 066404 (2006).

\bibitem{localization} C.L. Kane, P.A. Lee, N. Read, Phys. Rev. {\bf B 39}
6880 (1989).
G. Mart\'\i nez, P. Horsch, Phys. Rev. {\bf B 44}, 317 (1991).


\end{thebibliography}
\end{document}